# Zero Correlation Zone Sequences With Flexible Block-Repetitive Spectral Constraints

Branislav M. Popovic, Peng Wang, Fredrik Berggren and Renaud-Alexandre Pitaval

*Abstract*—A general construction of a set of time-domain sequences with sparse periodic correlation functions, having multiple segments of consecutive zero-values, i.e. multiple zero correlation zones (ZCZs), is presented. All such sequences have a common and block-repetitive structure of the positions of zeros in their Discrete Fourier Transform (DFT) sequences, where the exact positions of zeros in a DFT sequence do not impact the positions and sizes of ZCZs. This property offers completely new degree of flexibility in designing signals with good correlation properties under various spectral constraints. The non-zero values of the DFT sequences are determined by the corresponding frequency-domain modulation sequences, constructed as the element-by-element product of two component sequences: a "long" one, which is common to the set of time-domain sequences, and which controls the peak-to-average power ratio (PAPR) properties of the time-domain sequences; and a "short" one, periodically extended to match the length of the "long" component sequence, which controls the non-zero crosscorrelation values of all time-domain sequences. It is shown that 0 dB PAPR of time-domain sequences can be obtained if the "long" frequency-domain component sequence is selected to be a *modulatable* constant amplitude zero autocorrelation (MCAZAC) sequence. A generalized and simplified unified construction of MCAZAC sequences is presented.

## I. INTRODUCTION

THE design of spectrally constrained time-domain sequences with good correlation properties in essence relates to the design of their Discrete Fourier Transform (DFT) coefficients, under certain constraints on the positions of the non-zero values within each sequence of DFT coefficients (called DFT sequences in the sequel). Such spectral constraint has been described in terms of *forbidden* [1] or *unavailable* [2] frequencies of harmonically related complex exponentials (called carriers in [1] and subcarriers in [2]) that constitute the basis for the DFT decomposition of each time-domain sequence.

The two constructions presented in [1] produce the time domain sequences with uniformly low periodic autocorrelation function sidelobes for a special spectral constraint defined through a set of distributed spectral null frequencies whose positions within a DFT sequence form a cyclic difference set [1, Theorem 3]. Additionally, to achieve a constant envelope of the time-domain sequences, both of these constructions rely on the ideal periodic autocorrelation of the DFT sequences (which includes spectral nulls). Obviously, independent optimizations of the autocorrelation function and of the peak-to-average power ratio (PAPR) of the resulting time-domain sequences are not possible by these constructions, as both are determined by a common spectral constraint (spectral null frequencies forming a cyclic difference set).

The construction presented in [2] produces the sets of so-called *quasi* zero correlation zone (quasi-ZCZ) sequences for arbitrary spectral constraints. The pairwise crosscorrelation of these sequences is zero in a zone of delays around and including zero delay, but, as opposed to the conventional ZCZ sequences [3]-[10], the periodic out-of-phase autocorrelations are *not* zero in the same zone of delays. In some applications the non-zero autocorrelation sidelobes might cause inferior performances compared to those obtained by deploying sequences with zero autocorrelation zones [5].

The design of spectrally constrained ZCZ sequences has been considered also in [3, Definition 8], but only for a single and very specific class of spectral constraints, defined through "$L$-fold expander" operator, which converts a length-$N$ sequence $\{x(n)\}$ into a length-$LN$ DFT sequence $\{x_E(n)\}$ by inserting $L-1$ zeros after each element of the sequence.

A generalized version of the above class of spectral constraints is considered in [11] and [12, p. 42, Alt-1] for the design of ZCZ sequences, where identical and equidistant blocks of zeros are interlaced with the identical and equidistant blocks of non-zero DFT coefficients. A block of uniformly spaced subcarrier frequencies corresponding to a block of non-zero DFT coefficients has been labelled as a *physical resource block* (PRB) in [11]-[12], containing 12 subcarrier frequencies, where a set of PRBs in a given bandwidth has been denoted as an *interlace*. Equidistant spacing of PRBs produces time-domain sequences whose periodic autocorrelation functions have zero autocorrelation zones (ZAZ) of length inversely proportional to the frequency spacing between the PRBs. An irregular i.e. non-equidistant spacing of PRBs may result in the reduction of the periodic autocorrelation sidelobes of the corresponding time-domain sequences, although at the expense of losing zero autocorrelation zones [12, p.42, Alt-2].

A natural question is then whether is it possible to generate the sets of time-domain sequences with ZCZ in their pairwise periodic crosscorrelations and ZAZ in their periodic autocorrelation functions by using more flexible frequency allocations, where such flexible frequency allocations would have several purposes: to satisfy a range of different system spectral constraints, to control the maximum correlation values outside ZCZ and ZAZ, and to ensure low PAPR of the time-domain sequences. In this paper we show how to achieve that goal.

The paper is organized as follows. In Section II we give some basic definitions. General construction of an optimum set of ZCZ sequences sharing a common flexible spectral constraint is presented in Section III. The correlation properties of the sets of sequences obtained from the general construction are derived in Section IV. In Section V we present a general construction of frequency-domain modulation sequences producing 0 dB PAPR of spectrally constrained time-domain sequences. In Section VI we consider constructions of multiple sets of spectrally constrained ZCZ sequences with bounded crosscorrelations. Finally, a few concluding remarks are given in Section VII.

## II. BASIC DEFINITIONS

The DFT of a time-domain sequence $x'(k)$, $k = 0, 1, \dots, N-1$, produces the DFT sequence $X'(f)$, $f = 0, 1, \dots, N-1$, according to [13, p. 516]

$$X'(f) = \sum_{k=0}^{N-1} x'(k) W_N^{fk} \qquad (1)$$

where $W_N = \exp(-\sqrt{-1}\, 2\pi/N)$. The sequence $\{x'(k)\}$ can be reconstructed from the DFT sequence $\{X'(f)\}$ according to [13, p. 516]

$$x'(k) = \frac{1}{N} \sum_{f=0}^{N-1} X'(f) W_N^{-kf} \qquad (2)$$

It can be easily seen that the sequence (2) has $N$ times smaller energy than the sequence (1), i.e.

$$\sum_{k=0}^{N-1} |x'(k)|^2 = \frac{1}{N} \sum_{f=0}^{N-1} |X'(f)|^2. \qquad (3)$$

When the design of a time-domain sequence is performed by the design of a DFT sequence, as it is the case in this paper, for the evaluation purposes it is convenient that the time-domain sequence $\{x(k)\}$ has the same energy as its DFT sequence, i.e.

$$x(k) \triangleq \frac{1}{\sqrt{N}} \sum_{f=0}^{N-1} X(f) W_N^{-kf}. \qquad (4)$$

From (1) and (4) it follows that the DFT the time-domain sequence $\{x(k)\}$ produces the sequence $\{X(f)\}$ given by

$$X(f) = \frac{1}{\sqrt{N}} \sum_{k=0}^{N-1} x(k) W_N^{fk}. \qquad (5)$$

The periodic crosscorrelation function $\theta_{xy}(p)$ between the sequences $\{x(k)\}$ and $\{y(k)\}$ of length $N$ is usually defined as

$$\theta_{xy}(p) = \begin{cases} \sum_{k=0}^{N-1} x(k) y^*(k+p), & p \geq 0 \\ \theta_{yx}^*(-p), & p < 0 \end{cases} \qquad (6)$$

where $p$ is a cyclic shift of $\{y(k)\}$, and "*" denotes the complex conjugation. From (4) and (6) it follows that

$$\theta_{xy}(p) = \begin{cases} \sum_{f=0}^{N-1} X(f) Y^*(f) W_N^{pf}, & p \geq 0 \\ \theta_{yx}^*(-p), & p < 0 \end{cases} \qquad (7)$$

where $\{X(f)\}$ and $\{Y(f)\}$ are DFT sequences of $\{x(k)\}$ and $\{y(k)\}$ respectively, as given in (5).

The periodic autocorrelation function $\theta_{xx}(p)$ is said to have a Zero Autocorrelation Zone (ZAZ) of length $D_{ZAZ}$ if

$$\theta_{xx}(p) \begin{cases} = 0, & |p| = 1, 2, \dots, D_{ZAZ} \\ \neq 0, & |p| = D_{ZAZ} + 1 \end{cases}. \qquad (8)$$

The periodic crosscorrelation function $\theta_{xy}(p)$ is said to have a Zero Crosscorrelation Zone (ZCCZ) of length $D_{ZCCZ}$ if

$$\theta_{xy}(p) \begin{cases} = 0, & |p| = 0, 1, \dots, D_{ZCCZ} \\ \neq 0, & |p| = D_{ZCCZ} + 1 \end{cases}. \qquad (9)$$

The upper bound of $D_{ZCZ} = \min\{D_{ZCCZ}, D_{ZAZ}\}$ for the set of $M$ ZCZ sequences of length $N$ with arbitrary alphabet of complex numbers is given by [9]

$$D_{ZCZ} \leq \frac{N}{M} - 1. \qquad (10)$$

An optimum set of ZCZ sequences is the one having $D_{ZCZ}$ satisfying (10) with equality. Some special ZCZ sequences, such as those constructed in the next Section, have periodic correlation functions having multiple segments of consecutive zero-values. For those sequences we can define multiple ZAZ and ZCCZ lengths by applying (8) and (9) to appropriate cyclically shifted versions of these sequences.

## III. GENERAL CONSTRUCTION

The general construction of a set of spectrally-constrained time-domain sequences we are going to present in this Section consists of two key ingredients: A) specifying a *block-repetitive* structure of spectral constrains (or alternatively, allowed i.e. used frequencies); and B) specifying a set of two-component frequency-domain modulation sequences, whose elements modulate DFT subcarriers at allowed frequencies.

The sequence of DFT frequencies $f = 0, 1, \dots, N-1$, $N = \delta t$, where $\delta$ and $t$ are positive integers, is divided into $t$ *subbands* of $\delta$ consecutive and equidistant discrete frequencies. The same *subset* of $A$ ($A \leq \delta$) frequencies in each subband is selected to be used for transmission.

A set of $A$ spectrally-constrained time-domain sequences $\{s_n(k)\}$ is obtained from (4), from a corresponding set of DFT sequences $\{S_n(f)\}$, generated from a set of unit-magnitude frequency-domain modulation sequences $c_n(u)$, $n = 0, 1, \dots, A-1$, $u = 0, 1, \dots, L-1$, $L = At$, according to

$$s_n(k) \triangleq \frac{1}{\sqrt{N}} \sum_{f=0}^{N-1} S_n(f) W_N^{-kf}$$

$$S_n(f) = \begin{cases} c_n(Ai + l), & f = \delta i + j_l \in \bar{\Omega} \\ 0, & \text{otherwise} \end{cases}$$

$$i = 0, 1, \dots, t-1$$

$$0 \leq j_0 < j_1 < \cdots < j_{A-1} \leq \delta - 1 \qquad (11)$$

where $\bar{\Omega}$ is an *interlace* - a sequence of $L = At$ allowed frequencies.

Each of $A$ frequency-domain modulation sequences $\{c_n(u)\}$ is constructed as

$$c_n(u) = b_n(u \bmod A)\, a(u)$$

$$u = 0, 1, \dots, L-1, \quad L = At \qquad (12)$$

where $b_n(l)$, $l = 0, 1, \dots, A-1$, is the $n$-th ($n = 0, 1, \dots, A-1$) sequence of length $A$ in a set of $A$ unit-magnitude orthogonal sequences. The sequence $\{a(u)\}$ is a unit-magnitude sequence of length $L$.

An interesting special case of the sequences (12) is obtained when $\{a(u)\}$ is a Zadoff-Chu (ZC) sequence, defined as [14]

$$a(u) = W_{At}^{\alpha u(u+At \bmod 2+2q)/2} \tag{13}$$

where $\alpha$ is an integer, often called the *root-index*, relatively prime to $At$, where $q$ is also an integer, while the sequences $\{b_n(l)\}$ are defined as

$$b_n(u \bmod A) = a(tn) W_A^{\alpha n(u \bmod A)}. \tag{14}$$

By inserting (13) and (14) into (12) we obtain

$$c_n(u) = a(u)a(tn) W_{At}^{\alpha utn}. \tag{15}$$

As any cyclically-shifted version $a(u + p)$ of a ZC sequence can be decomposed as [14]

$$a(u + p) = a(u)a(p) W_{At}^{\alpha up} \tag{16}$$

we obtain

$$c_n(u) = a(u + tn) \tag{17}$$

by applying (16) to (15).

## IV. CORRELATION PROPERTIES

In this Section we will derive some properties of the periodic correlation functions of the time-domain sequences $\{s_n(k)\}$.

### A. Periodic Crosscorrelation

From (7), (11) and (12) we obtain

$$\theta_{s_x s_y}(p \geq 0) = \sum_{l=0}^{A-1} b_x(l) b_y^*(l) W_N^{p j_l} \sum_{i=0}^{t-1} W_t^{pi}. \tag{18}$$

The inner sum in (18) is zero for $p \not\equiv 0 \pmod t$. For $p = tq$, $q = 0,1,\dots,\delta-1$, from (18) we obtain

$$\theta_{s_x s_y}(qt) = t \sum_{l=0}^{A-1} b_x(l) b_y^*(l) W_\delta^{q j_l}. \tag{19}$$

As $\theta_{s_x s_y}(0) = 0$ (as $\{b_x(l)\}$ and $\{b_y(l)\}$ are orthogonal), from (19) we could conclude that the periodic crosscorrelation function has a ZCCZ of length at least $D_{ZCCZ} = t - 1$, which is the case when $\sum_{l=0}^{A-1} b_x(l) b_y^*(l) W_\delta^{j_l} \neq 0$. Thus we have

$$\theta_{s_x s_y}(p \geq 0) = \begin{cases} 0, & p = 0 \\ 0, & p \not\equiv 0 \pmod t \\ t \sum_{l=0}^{A-1} b_x(l) b_y^*(l) W_\delta^{q j_l}, & p = tq \\ & q = 1,\dots,\delta-1 \end{cases} \tag{20}$$

### B. Periodic Autocorrelation

From (18) we obtain

$$\theta_{s_n s_n}(p \geq 0) = \begin{cases} L, & p = 0 \\ 0, & p \not\equiv 0 \pmod t \\ t \sum_{l=0}^{A-1} W_\delta^{q j_l}, & p = tq, \ q = 1,\dots,\delta-1 \end{cases} \tag{21}$$

From (21) we could conclude that the periodic autocorrelation function has a ZAZ of length *at least* $D_{ZAZ} = t - 1$, which is the case when $\sum_{l=0}^{A-1} W_\delta^{j_l} \neq 0$. However, this sum could be equal to zero in some special cases of $j_l$. For example, if both $A$ and $\delta$ can be decomposed as products of some common integers, i.e. if $A = A'\sigma$ and $\delta = A'\sigma B$, then we can define $j_l$ as

$$j_l = \frac{\delta}{\sigma} i' + j'_{l'}$$

$$0 \leq j'_0 < j'_1 < \cdots < j'_{A'-1} \leq \frac{\delta}{\sigma} - 1$$

$$l = A' i' + l', \ i' = 0, 1, \dots, \sigma - 1, \ l' = 0, 1, \dots, A' - 1. \tag{22}$$

Then it follows from (21) and (22) that

$$\sum_{l=0}^{A-1} W_\delta^{q j_l} = \sum_{l'=0}^{A'-1} W_\delta^{q j'_{l'}} \sum_{i'=0}^{\sigma-1} W_\sigma^{q i'}$$

$$= 0, \quad \text{for } q \not\equiv 0 \pmod \sigma \text{ i.e. } p \not\equiv 0 \pmod{t\sigma}. \tag{23}$$

From (23) we see that the periodic autocorrelation function has a ZAZ of length $D_{ZAZ} = t\sigma - 1$ if (22) holds, even if the short sequences $\{b_n(l)\}$ are of length $A$.

From (21) we can also see that when $A = \delta$, i.e. when the spectral constraint is removed, the sum $\sum_{l=0}^{A-1} W_\delta^{q j_l} = 0$ for any $q$, so the periodic autocorrelation becomes ideal.

The periodic correlation functions (20) and (21) can be described as *sparse*, because they have very few non-zero values, which can be controlled through the selection of allowed frequencies $\{j_l\}$.

## V. FREQUENCY-DOMAIN MODULATION SEQUENCES PRODUCING 0 DB PAPR OF SPECTRALLY CONSTRAINED TIME-DOMAIN SEQUENCES

In this Section it will be shown that the 0 dB PAPR of the time-domain sequences $\{s_n(k)\}$ can be achieved if the sequence $\{a(u)\}$ is a *modulatable* constant-amplitude zero-autocorrelation (MCAZAC) sequence.

### A. Modulatable CAZAC Sequences

The ZC sequence (13) can be classified as a CAZAC sequence. A CAZAC sequence $a(u), u = 0,1,\dots,L-1$ is characterized by two properties: it has a constant magnitude, and it has the *ideal* periodic autocorrelation function $\theta_{aa}(p) = 0, |p| = 1,2,\dots,L-1$. As a consequence of these two properties, the DFT of a CAZAC sequence $\{a(u)\}$ produces the sequence $\{\Lambda(z)\}$ that is also a CAZAC sequence (often called a *Fourier dual* of $\{a(u)\}$ if it has the same magnitude as $\{a(u)\}$ [15]). To show that $\{\Lambda(z)\}$ is indeed a CAZAC sequence, let us assume without loss of generality that $\{\Lambda(z)\}$ is obtained from (5), so from (4) we can write

$$a(u) = \frac{1}{\sqrt{L}} \sum_{z=0}^{L-1} \Lambda(z) W_L^{-uz}. \tag{24}$$

Then from (7) we can write $\theta_{aa}(p) = \sum_{z=0}^{N-1} |\Lambda(z)|^2 W_N^{pz}$ for $|p| \geq 0$, where we see that $\theta_{aa}(p) = 0$ for $|p| > 0$ if and only if $|\Lambda(z)|$ is a constant.

It remains to show that the sequence $\{\Lambda(z)\}$ has the *ideal* periodic autocorrelation function $\theta_{\Lambda\Lambda}(p)$. If we introduce in (24) the change of variables $u = L - v$, $v = 0, 1, \ldots, L - 1$, where the $a(L) = a(0)$ due to the periodicity of the DFT, we obtain

$$a(L - v) = \frac{1}{\sqrt{L}} \sum_{z=0}^{L-1} \Lambda(z) W_L^{vz}. \quad (25)$$

The right-hand side of (25) is the DFT of the sequence $\{\Lambda(z)\}$, while the left-hand side of (25) is reversed and cyclically shifted version of the sequence $\{a(u)\}$. Thus from (7) and (25) we obtain

$$\theta_{\Lambda\Lambda}(p) = \sum_{v=0}^{N-1} |a(L - v)|^2 W_N^{pv} \quad (26)$$

which is zero for $|p| > 0$ because the sequence $\{a(u)\}$ by definition has a constant magnitude.

A CAZAC sequence $\{a(u)\}$ is called *modulatable* if it can be represented as a product of two sequences, a specific carrier sequence $\{\chi(u)\}$ of length $L = At$ and an *arbitrary*, shorter and $t$ times periodically extended, modulation sequence $\{\eta(l)\}$ of length $A$, i.e. when

$$a(u) = \chi(u)\eta(u \bmod A), \quad u = 0, 1, \ldots, L - 1, L = At. \quad (27)$$

The corresponding DFT sequence $\{\Lambda(z)\}$ of the same magnitude can be obtained from (5) as

$$\Lambda(z) = \frac{1}{\sqrt{L}} \sum_{u=0}^{L-1} a(u) W_L^{zu}$$

$$= \frac{1}{\sqrt{L}} \sum_{l=0}^{A-1} \eta(l) W_L^{zl} \sum_{i=0}^{t-1} \chi(Ai + l) W_t^{zi} \quad (28)$$

where we used the change of variables $u = Ai + l$.

To have $|\Lambda(z)| = constant$ for $0 \leq z < L$ and for *arbitrary* constant-magnitude sequence $\{\eta(l)\}$, the sum over $l$ must have only one non-zero element for any $z$. This is possible if and only if the sum over $i$ is non-zero and has the absolute value equal to $|\Lambda(z)|$ only for a single value of $l$ for given value of $z$. If the MCAZAC sequence $\{a(k)\}$ has the unit magnitude, then from (28) it follows that the non-zero absolute value of the sum over $i$ is equal to $\sqrt{L}$. Thus we have

$$\left|\sum_{i=0}^{t-1} \chi(Ai + l) W_t^{zi}\right| = \begin{cases} \sqrt{L}, & l = \hat{l} \\ 0, & l \neq \hat{l}. \end{cases} \quad (29)$$

As we can write

$$\left|\sum_{i=0}^{t-1} \chi(Ai + l) W_t^{zi}\right| = \left|\frac{1}{\eta(l)}\right| \left|\sum_{i=0}^{t-1} a(Ai + l) W_t^{zi}\right| \quad (30)$$

from (29) and (30) we obtain the following Lemma.

*Lemma*: A sequence $\{a(k)\}$ is a MCAZAC sequence (27) of the *unit* magnitude, iff

$$\left|\sum_{i=0}^{t-1} a(Ai + l) W_t^{zi}\right| = \begin{cases} \sqrt{L}, & l = \hat{l} \\ 0, & l \neq \hat{l} \end{cases} \quad (31)$$

where $\hat{l}$ is one and only one value of $l$ among $\{0, 1, \ldots, A - 1\}$ for given value of $z$.

*B. Generalized Unified Construction of Modulatable CAZAC Sequences*

All known MCAZAC sequences are of length that can be expressed as an integer multiple of another squared integer. It means that for given $t$ we can select any $A$ that is a factor of $t$, i.e. $t = sA$, and generate the corresponding MCAZAC sequence $\{a(k)\}$ of length $L = At = sA^2$.

The earliest construction of MCAZAC construction has been published in [16], producing modulatable Frank sequences of length $L = A^2$. The MCAZAC sequences of length $L = sA^2$ have been constructed in [17], as a class of generalized chirp-like (GCL) sequences obtained by modulating ZC sequences of length $L = sA^2$, i.e. by using (27) where the carrier sequence $\{\chi(k)\}$ is a ZC sequence of length $L = sA^2$. GCL sequences with minimum alphabets have been derived in [18], where it has been shown that the modulatable Frank sequences are a special case of GCL sequences. The GCL sequences are further generalized in [19], in the so-called unified construction of MCAZAC sequences, given by

$$a(iA + l) = \eta(l)\, g_l(i \bmod s) W_t^{r_1 \mu(l) + n_1}$$

$$g_l(k) = W_s^{\phi_l k^2 / 2}, \quad \phi_l = (s + 1)\left[r_0 + n_0 \frac{l(l+1)}{2}\right]$$

$$(s + 1)n_0 \equiv 0 \pmod{2}, \quad \left(r_0 + n_0 \frac{l(l+1)}{2}, s\right) = 1, (r_1, A) = 1$$

$$i = 0, 1, \ldots, t - 1, \; l = 0, 1, \ldots, A - 1, \; k = 0, 1, \ldots, s - 1$$

$$(32)$$

where $t = sA$, $\eta(l)$ is an arbitrary unit-magnitude complex number, $r_0$, $n_0$, $r_1$ and $n_1$ are integers, and $\mu$ is a permutation over the set $\{0, 1, \ldots, A - 1\}$.

The term $[r_0 + n_0 l(l + 1)/2]$ in the unified construction (32) produces a set of integers with a specific structure, but with the only requirement that all these integers are relatively prime to $s$. These integers are used as the exponents of $W_2$ and $W_{2s}$, where they are effectively reduced $\pmod{2}$ and $\pmod{2s}$, respectively, hinting that *any* corresponding set of integers relatively prime to $s$ would also produce the MCAZAC sequences. It means that the above mentioned term can be replaced with a single coefficient $r_0(l)$, being an integer that can be independently selected for each $l$, with the only condition that $(r_0(l), s) = 1$. Obviously, once the coefficient $n_0$ has been removed we could also remove the condition $(s + 1)n_0 \equiv 0 \pmod{2}$. Thus we can generalize $g_l(k)$ in (32) as

$$g_l(k) = W_s^{\phi_l k^2 / 2}, \quad \phi_l = (s + 1)r_0(l), \quad (r_0(l), s) = 1 \quad (33)$$

If $s$ is even, the sequence $\{g_l(k)\}$ in (33) is a ZC sequence that can be obtained from (13) with the root index $\alpha = \phi_l$ and $q = 0$. If $s$ is odd, it can be represented as $s = 2\psi - 1$, so we have

$$g_l(k) = W_s^{\psi r_0(l) k^2} \quad (34)$$

where $\psi r_0(l)$ is relatively prime to $s$, as both $\psi$ and $r_0(l)$ are relatively prime to $s$. The sequence $\{g_l(k)\}$ in (34) can be obtained from (13) by applying the identity

$$\frac{1}{2} \equiv \psi \pmod{s} \quad (35)$$

and by selecting $\alpha = r_0(l)$ and $q = -\psi$, so it is a ZC sequence. Therefore one obvious way to generalize (32) is to define $\{g_l(k)\}$ as *any* ZC sequence of length $s$.

An additional direction of generalization of (32) is to simplify the term $[r_1\mu(l) + n_1](\mod t)$. For each $l = 0,1, \dots, A - 1$, the term $[r_1\mu(l) + n_1](\mod t)$ can be expressed as

$$[r_1\mu(l) + n_1] \mod t = \hat{\mu}(l) + m_l A$$
$$\hat{\mu}(l) = \{[r_1\mu(l) + n_1] \mod t\} \mod A$$
$$m_l = \left\lfloor \frac{[r_1\mu(l)+n_1] \mod t}{A} \right\rfloor \quad (36)$$

where we used a general identity $x = x \mod y + \left\lfloor \frac{x}{y} \right\rfloor y$. Then we have

$$\hat{\mu}(l) = \{[r_1\mu(l) + n_1] \mod t\} \mod A$$
$$= \left\{ [r_1\mu(l) + n_1] - \left\lfloor \frac{r_1\mu(l)+n_1}{sA} \right\rfloor sA \right\} \mod A$$
$$= [r_1\mu(l) + n_1] \pmod{A}. \quad (37)$$

From (37) we see that $\hat{\mu}$ is a permutation over the set $\{0, 1, \cdots, A-1\}$ because due to $(r_1, A) = 1$ the product $r_1\mu(l) \mod A$ produces once each of the values from the set $\{0, 1, \cdots, A-1\}$, while $n_1$ shifts mod $A$ all these values by the same amount and thus preserves their diversity.

By substituting (36) into (32), we obtain

$$a(iA + l) = \eta(l)\hat{g}_l(i \mod s)W_t^{\hat{\mu}(l)i}$$
$$\hat{g}_l(i \mod s) = g_l(i \mod s)W_s^{m_l(i \mod s)} \quad (38)$$

where $\hat{g}_l(i \mod s)$ is another ZC sequence, because it is a linear-phase rotated version of the ZC sequence $g_l(i \mod s)$.

As the ZC sequences are CAZAC sequences, an immediate question is whether $\hat{g}_l(k)$ in (38) could be *any* unit-magnitude CAZAC sequence of length $s$, $s > 1$. The answer is positive, as it will be proven soon, so we can now define a *new generalized unified construction of MCAZAC sequences* as

$$a(iA + l) = \eta(l)g_l(i \mod s)W_t^{\mu(l)i}$$
$$i = 0,1, \dots, t-1, \ l = 0,1, \dots, A-1, \ k = 0,1, \dots, s-1 \quad (39)$$

where $t = sA$, $\eta(l)$ is an arbitrary unit-magnitude complex number, $g_l(k)$ is either the $k$-th element of a unit-magnitude CAZAC sequence of length $s$ (if $s > 1$) or 1 (if $s = 1$), and $\mu$ is a permutation over the set $\{0, 1, \dots, A-1\}$.

To prove that (39) produces an MCAZAC sequence $\{a(u)\}$ it is sufficient to show that such a sequence satisfies Lemma (31). If the sum in the left-hand side of (31) is denoted as $S(l) = \sum_{i=0}^{t-1} a(Ai + l)W_t^{zi}$, then it follows that

$$S(l) = \eta(l)\sum_{i=0}^{t-1}g_l(i \mod s)W_t^{[\mu(l)+z]i} \quad (40)$$

If we introduce the change of variables

$$i = js + k, \quad j = 0, 1, \dots, A-1, \ k = 0, 1, \dots, s-1 \quad (41)$$

we obtain

$$S(l) = \eta(l)\sum_{j=0}^{A-1}W_A^{[\mu(l)+z]j}$$
$$\sum_{k=0}^{s-1}g_l(k)W_t^{[\mu(l)+z]k}. \quad (42)$$

As $\mu$ is a permutation over the set $\{0, 1, \dots, A-1\}$, the sum over $j$ is non-zero only for a single value of $\mu(\hat{l})$ such that

$$\mu(\hat{l}) + z \equiv 0 \pmod{A} \quad (43)$$

where $\hat{l}$ is one and only one value of $l$ among $\{0,1, \dots, A-1\}$ for given value of $z$. From (43) we can write that

$$\mu(\hat{l}) + z = xA \quad (44)$$

where $x$ is an integer. From (42) and (44) we obtain

$$S(\hat{l}) = \eta(\hat{l})A\sum_{k=0}^{s-1}g_l(k)W_s^{xk}. \quad (45)$$

The sum over $k$ in (45) is a DFT coefficient of a sequence $\{g_l(k)\}$ of length $s$. If $s = 1$, obviously $|S(\hat{l})| = A = \sqrt{L}$. If $s > 1$, $\{g_l(k)\}$ is a CAZAC sequence of unit magnitude. From (3) it follows that the DFT coefficients of such CAZAC sequence have the magnitude equal to $\sqrt{s}$. This leads to $|S(\hat{l})| = A\sqrt{s} = \sqrt{L}$, which concludes the proof.

### C. 0 dB PAPR

We will use now Lemma (31) to prove that the time-domain sequences $\{s_n(k)\}$, for arbitrary $\delta$ and arbitrary number $A \leq \delta$, have 0 dB PAPR, i.e. a constant magnitude, as long as the sequence $\{a(u)\}$ of length $L = At$ is an MCAZAC sequence, where $t$ is a multiple (including 1) of $A$. To prove that, we express the time-domain sequence $\{s_n(k)\}$ in (11) by using (12) as

$$s_n(k) = \frac{1}{\sqrt{N}}\sum_{l=0}^{A-1}W_N^{-kjl}b_n(l)\sum_{i=0}^{t-1}W_t^{-ki}a(iA + l)$$
$$k = 0,1, \dots, N-1. \quad (46)$$

According to Lemma (31), if the sequence $\{a(k)\}$ is an MCAZAC sequence of the *unit* magnitude, then the sum over $i$ is non-zero and has the absolute value equal to $\sqrt{L} = \sqrt{At}$ only for a single value of $l$ for given value of $k$. Therefore we obtain

$$|s_n(k)| = \sqrt{A/\delta}, \quad k = 0,1, \dots, N-1 \quad (47)$$

which concludes the proof.

## VI. Multiple Sets of ZCZ Sequences With Bounded Crosscorrelations

In some applications the number of sequences in a single set of ZCZ sequences (11)-(12) might not be enough. One way to generate multiple sets of such ZCZ sequences over the same spectral constraints is to use multiple sets of $\{b_n(l)\}$ sequences. In that case the crosscorrelation between the two ZCZ sequences obtained from two different sets of $\{b_n(l)\}$ sequences is given by (20), except for the zero delay, in which case it is equal to $t\sum_{l=0}^{A-1}b_x(l)b_y^*(l)$.

Another way is to use multiple sequences $\{a(u)\}$. For example, starting from (39), we can define a set $\{a_r(u)\}$, $r = 0, 1, \dots, R-1$, such that

$$a_r(iA + l) = \eta(l)g_l(i \bmod s)W_t^{\mu_r(l)i}$$
$$i = 0,1,...,t-1, \ l = 0,1,...,A-1, \ k = 0,1,...,s-1 \quad (48)$$

where $t = sA$, $\eta(l)$ is an arbitrary unit-magnitude complex number, and $g_l(k)$ is either the $k$–th element of a unit-magnitude CAZAC sequence of length $s$ (if $s > 1$) or 1 (if $s = 1$); $\mu_r$ is a permutation from a set of permutations of the sequence $\{0, 1, ..., A-1\}$ such that for any two permutations $\mu_\alpha$ and $\mu_\beta$ it holds that

$$\mu_\alpha(l) - \mu_\beta(l) \equiv \mu_\gamma(l) \pmod{A} \quad (49)$$

where $\mu_\gamma(l)$ is the $l$-th element of a permutation $\mu_\gamma$ of the sequence $\{0, 1, ..., A-1\}$. For example, if $A$ is a prime number, a set of $A-1$ permutations satisfying (49) can be constructed as

$$\mu_r(l) \equiv rl \pmod{A}$$
$$r = 1, 2, ..., A-1, \quad l = 0, 1, ..., A-1. \quad (50)$$

The crosscorrelation between the two time-domain sequences $\{s_n(k)\}$, obtained from two different $\{a_\alpha(u)\}$ and $\{a_\beta(u)\}$ sequences and two different $\{b_x(l)\}$ and $\{b_y(l)\}$ sequences, is bounded according to

$$\left|\theta_{s_{(\alpha,x)}s_{(\beta,y)}}(p)\right| \leq t. \quad (51)$$

To prove (51), we start by combining (7), (48) and (49), to obtain

$$\theta_{s_{(\alpha,x)}s_{(\beta,y)}}(p) = \sum_{l=0}^{A-1} b_x(l)b_y^*(l)W_N^{pj_l}$$
$$\cdot \sum_{i=0}^{t-1} W_t^{[\Phi(l)+p]i} \quad (52)$$

where

$$\Phi(l) \equiv [\mu_\alpha(l) - \mu_\beta(l)] \pmod{t}, \ l = 0,1,...,A-1. \quad (53)$$

We will consider now how many non-zero terms has the sum over $i$ in (52), as function of $s$ and $l$.

If $s = 1$, $\Phi(l) = \mu_\gamma(l)$ is yet another permutation of the sequence $\{0, 1, ..., A-1\}$, so for any $p$ there will be only a single value $l = \chi_p$ such that $\Phi(\chi_p) \equiv -p \pmod{t}$. Thus we have

$$\theta_{s_{(\alpha,x)}s_{(\beta,y)}}(p) = tb_x(\chi_p)b_y^*(\chi_p)W_N^{pj_{\chi_p}}. \quad (54)$$

From (54) it follows that for $s = 1$ the maximum crosscorrelation magnitude between the two time-domain sequences of length $A^2$ belonging to two different ZCZ sets defined by the permutations $\mu_\alpha$ and $\mu_\beta$ is

$$\left|\theta_{s_{(\alpha,x)}s_{(\beta,y)}}(p)\right| = t \quad (55)$$

i.e. it is $A$ times lower than the energy of each time-domain sequence.

If $s > 1$, $\Phi(l)$ is not anymore a permutation of the sequence $\{0, 1, ..., A-1\}$, but still each value of $\Phi(l)$ appears only once when $l = 0, 1, ..., A-1$. To prove that we start by re-writing (49) as

$$\mu_\alpha(l) - \mu_\beta(l) = x_l A + \mu_\gamma(l) \quad (56)$$

where $x_l$ is an integer and $\mu_\gamma(l)$ is a third permutation of the sequence $\{0, 1, ..., A-1\}$. By substituting (56) into (53) we obtain

$$\Phi(l) \equiv x_l A + \mu_\gamma(l) \pmod{sA}. \quad (57)$$

Then for any two different values $l_0$ and $l_1$ from $\{0, 1, \cdots, A-1\}$ we have

$$\Phi(l_0) - \Phi(l_1) \equiv (x_{l_0} - x_{l_1})A$$
$$+ [\mu_\gamma(l_0) - \mu_\gamma(l_1)] \pmod{sA}. \quad (58)$$

Since $\mu_\gamma(l)$ is a permutation of the sequence $\{0, 1, ..., A-1\}$, we have

$$\mu_\gamma(l_0) - \mu_\gamma(l_1) \not\equiv 0 \pmod{A} \quad (59)$$

which leads to

$$(x_{l_0} - x_{l_1})A + [\mu_\gamma(l_0) - \mu_\gamma(l_1)] = \bar{x}_{l_0,l_1}A + \Delta_{l_0,l_1} \quad (60)$$

where $\bar{x}_{l_0,l_1}$ is an integer and $\Delta_{l_0,l_1}$ satisfies $0 < \Delta_{l_0,l_1} < A$. The integer $\bar{x}_{l_0,l_1}$ can be decomposed as a function of $s$ as

$$\bar{x}_{l_0,l_1} = \bar{x}_{l_0,l_1}^{(I)} s + \bar{x}_{l_0,l_1}^{(R)}$$
$$\bar{x}_{l_0,l_1}^{(I)} = \left\lfloor \frac{\bar{x}_{l_0,l_1}}{s} \right\rfloor, \ \bar{x}_{l_0,l_1}^{(R)} = \bar{x}_{l_0,l_1} \pmod{s} \quad (61)$$

From (60) and (61) we have

$$(x_{l_0} - x_{l_0})A + [\mu_\gamma(l_0) - \mu_\gamma(l_1)]$$
$$= \bar{x}_{l_0,l_1}^{(I)} sA + \bar{x}_{l_0,l_1}^{(R)} A + \Delta_{l_0,l_1}. \quad (62)$$

Finally, from (58) and (62) we obtain that

$$\Phi(l_0) - \Phi(l_1) \equiv \bar{x}_{l_0,l_1}^{(R)} A + \Delta_{l_0,l_1} \not\equiv 0 \pmod{sA} \quad (63)$$

because $0 \leq \bar{x}_{l_0,l_1}^{(R)} < s$ and $0 < \Delta_{l_0,l_1} < A$. That concludes the proof that each value of $\Phi(l)$ appears only once when $l = 0, 1, ..., A-1$.

Now we can go back to the consideration on how many non-zero terms has the sum over $i$ in (52), as a function of $l$ for $s > 1$. From (63) it follows that there will be $A$ different values of $\Phi(l)$, so it may happen that there is not a value $l = \chi_p$ such that $\Phi(\chi_p) \equiv -p \pmod{sA}$, in which case $\theta_{s_{(\alpha,x)}s_{(\beta,y)}}(p)$ would be zero. This case is reflected in the inequality part of (51), and it concludes the proof of it.

## VII. CONCLUDING REMARKS

A general class of spectrally constrained time-domain sequences, offering very flexible adaptation to different system requirements while preserving sparse correlation functions with multiple zero correlation zones, has been presented. As a side-product, a generalized and simplified version of the unified construction of modulatable CAZAC sequences has been derived.